\documentclass[letter,twocolumn]{jpsj2} 
\newcommand{\Romannum}[1]{\uppercase\expandafter{\romannumeral#1}}
\usepackage{bm}
\usepackage{mathrsfs}
\usepackage{exscale}
\usepackage{color}

\title{Evolution from Itinerant Antiferromagnet to Unconventional Superconductor \\
with Fluorine Doping in La(O$_{1-x}$F$_{x}$)FeAs\\ 
Revealed by $^{75}$As and $^{139}$ La Nuclear Magnetic Resonance}
\author{
Yusuke \textsc{Nakai}$^{1}$\thanks{E-mail address: nakai@scphys.kyoto-u.ac.jp},
Kenji \textsc{Ishida}$^{1}$\thanks{E-mail address: kishida@scphys.kyoto-u.ac.jp},
Yoichi \textsc{Kamihara,}$^{2}$
Masahiro \textsc{Hirano,}$^{2, 3}$\\
and Hideo \textsc{Hosono}$^{2, 3, 4}$}
\inst{$^{1}$Department of Physics, Graduate School of Science, Kyoto University, Kyoto 606-8502, Japan, \\
$^{2}$ERATO-SORST, JST, Frontier Collaborative Research Center, Tokyo Institute of Technology, Yokohama 226-8503, Japan,\\
$^{3}$Frontier Research Center, Tokyo Institute of Technology, Yokohama 226-8503, Japan,\\
$^{4}$Materials and Structures Laboratory, Tokyo Institute of Technology, Yokohama 226-8503, Japan\\}

\abst{We report experimental results of $^{75}$As and $^{139}$La nuclear magnetic resonance (NMR) in the iron-based layered La(O$_{1-x}$F$_{x}$)FeAs ($x=$ 0.0, 0.04 and 0.11). 
In the undoped LaOFeAs, $1/T_1$ of $^{139}$La exhibits a distinct peak at $T_N \sim$ 142 K below which the spectra become broadened due to the internal magnetic field attributed to an antiferromagnetic (AFM) ordering. 
In the 4\% F-doped sample, $1/T_1T$ exhibits a Curie-Weiss temperature dependence down to 30 K, suggesting the development of AFM spin fluctuations with decreasing temperature.
In the 11\% F-doped sample, in contrast, pseudogap behavior is observed in $1/T_1T$ both at the $^{75}$As and $^{139}$La site with a gap value of $\Delta_{\rm PG}\sim 172$ K. 
The spin dynamics vary markedly with F doping, which is ascribed to the Fermi-surface structure.  
As for the superconducting properties for the 4\% and 11\% F-doped samples, $1/T_1$ in both compounds does not exhibit a coherence peak just below $T_c$ and follows a $T^3$ dependence at low temperatures, which suggests unconventional superconductivity with line-nodes. 
We discuss similarities and differences between La(O$_{1-x}$F$_{x}$)FeAs and cuprates, and also discuss the relationship between spin dynamics and superconductivity on the basis of F doping dependence of $T_c$ and $1/T_1$. }

\kword{LaOFeAs, superconductivity, NMR}

\begin{document}
\maketitle
%
The recent discovery of iron-based layered superconductor La(O$_{1-x}$F$_x$)FeAs has opened up a new route to high-temperature superconductivity\cite{KamiharaFeAs}, because it has been reported that superconducting transition temperature $T_c$ of $R$(O$_{1-x}$F$_x$)FeAs ($R$: rare-earth) raises up to $\sim$ 50 K, which is the highest other than high-$T_c$ cuprates\cite{XHChenSm, GFChenCe, ZhiRenNd}. The crystal structure of La(O$_{1-x}$F$_x$)FeAs is tetragonal ($P4/nmm$), and consists of the LaO and FeAs layers which are stacked along the $c$ axis. 
Therefore, the physical properties are considered to be highly two dimensional, similar to the cuprate, ruthenate and cobaltate superconductors\cite{Singh}. Comparing between the Fe-based and cuprate superconductors, we notice their similarities and differences. 
One of similarities is that superconductivity of the two compounds is induced when carriers are introduced by means of elemental substitutions. 
In La(O$_{1-x}$F$_x$)FeAs, superconductivity emerges when 4\% F is doped to LaOFeAs. On the other hand, one of differences is that the parent compound LaOFeAs is metallic in contrast to the Mott insulator La$_2$CuO$_4$, although LaOFeAs shows AFM ordering around 150 K\cite{KamiharaFeAs, Dong, Cruz, McGuire, Nomura}. Other difference is that $T_c$ seems to be insensitive to $x$, i.e, $T_c$ remains almost unchanged from $x$ = 0.04 to 0.11\cite{KamiharaFeAs}. 
The F-concentration dependence of $T_c$ is in contrast with the ``dome dependence'' of $T_c$ in the hole-doped cuprates.

In order to shed light on magnetic properties in the undoped LaOFeAs and on their evolution with F doping as well as superconducting (SC) properties, we have performed La- and As-NMR measurements in La(O$_{1-x}$F$_x$)FeAs. In this paper, we mainly report temperature dependence of spin-lattice relaxation rate $1/T_1$, which is related to the wave vector $q$-averaged dynamical susceptibility. 
We show that LaOFeAs is an itinerant antiferromagnet with $T_N \sim 142$ K from the $T$-dependence of $1/T_1$. The magnetic ordering is suppressed by F doping, and is not observed in the SC samples. The magnetic fluctuations revealed by $1/T_1$ of As vary significantly with F doping, although $T_c$ is insensitive to F concentration. $1/T_1$ in the SC samples suggests that La(O$_{1-x}$F$_x$)FeAs is an unconventional superconductor with line-nodes in the SC gap. We consider that competition between itinerant antiferromagnetism and unconventional superconductivity also exists in the La(O$_{1-x}$F$_x$)FeAs system, as in unconventional superconductors realized in strongly correlated electron systems.          

Polycrystalline samples of La(O$_{1-x}$F$_{x}$)FeAs ($x=$0, 0.04 and 0.11) synthesized by solid-state reactions\cite{KamiharaFeAs} are ground into powder for NMR measurements. 
$T_c$ for $x=$0.04 and 0.11 determined from the onset of the SC transition with ac susceptibility is 17.5 K and 22.7 K, respectively.

\begin{figure}[tb]
\begin{center}
\includegraphics[width=6.8cm]{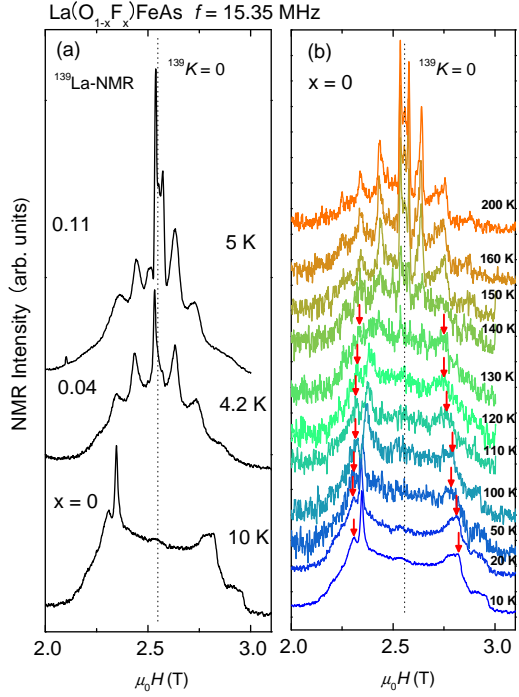}
\end{center}
\caption{(a) Field-sweep $^{139}$La NMR spectra at 15.35 MHz for $x=$0, 0.04, and 0.11 in La(O$_{1-x}$F$_x$)FeAs. (b) $^{139}$La-NMR spectra at a series of temperatures through $T_N\sim142$ K in the undoped LaOFeAs. The spectra have been offset vertically. Arrows indicate the fields at which we measured the internal field.}
\end{figure}
Figure~1(a) shows $^{139}$La($I=7/2$)-NMR spectra for $x=$0, 0.04, and 0.11 in La(O$_{1-x}$F$_x$)FeAs obtained by sweeping magnetic field at a fixed frequency 15.35 MHz. 
In the 4\% and 11\% F-doped samples which exhibit superconductivity, typical powder-pattern NMR spectra broadened by the first-order electric-quadrupole interaction are observed, which consist of seven peaks. 
The powder pattern La-NMR spectra indicate the absence of the internal fields at the La site. In contrast, quite different La-NMR spectrum is observed in the undoped LaOFeAs at 10 K; the central peak arising from the $1/2 \leftrightarrow -1/2$ transition is broadened due to the internal field. 
With increasing temperature, the broadening of the spectrum denoted by two red arrows decreases gradually and typical powder-pattern spectra similar to those observed for $x=$0.04 and 0.11 are obtained above 150 K. 
Along with a little anomaly of the susceptibility around 150 K\cite{KamiharaFeAs}, the observed spectrum for the undoped LaOFeAs suggests an AFM ordering occurs below $\sim150$ K. 
The internal field at the La site arises from the hyperfine field from Fe atoms, and is related with the Fe ordered moments. 
\begin{figure}[tb]
\begin{center}
\includegraphics[width=7cm]{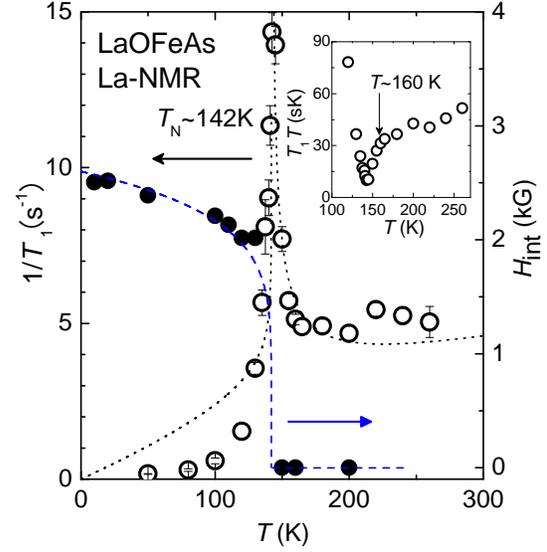}
\end{center}
\caption{$T$-dependence of $1/T_1$ ($\circ$) and the internal field ($\bullet$) of LaOFeAs. The dotted line is a fit to the SCR theory for weak itinerant antiferromagnet. The inset shows a plot of $T_1T$ as a function of $T$, indicating an abrupt drop around $\sim160$ K.}
\end{figure}
\begin{figure}[tb]
\begin{center}
\includegraphics[width=6.3cm]{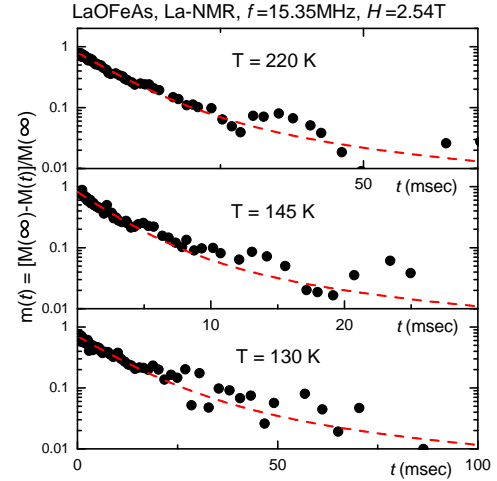}
\end{center}
\caption{The recoveries of the nuclear magnetization of La in LaOFeAs fitted uniquely with $m(t)=0.0119e^{(-t/T_1)}+0.0682e^{(-6t/T_1)}
+0.2061e^{(-15t/T_1)}+0.7137e^{(-28t/T_1)}$ (dotted lines).}
\end{figure}

The internal field $H_{\rm int}$ determined from the half width of the broadening is shown in Fig.~2 (filled circle). The dashed line in Fig.~2 is a fit to $M(T)/M_0 = [1-(T/T_N)]^{0.15}$. The growth of $M(T)/M_0$ below $T_N$ (=142 K, see below) is much steeper than 3-D mean field value (0.5), and the transition to the AFM state is almost first-order like. 
We consider that two-dimensionality of magnetic fluctuations is responsible for the reduced value of the critical exponent because similar rapid growth was reported in the cuprate antiferromagnet La$_2$CuO$_4$\cite{MacLaughlinLa2CO4}.

Further insight into the magnetic properties in LaOFeAs is obtained from $1/T_1$ measurements. 
$T_1$ was measured at the fixed magnetic field of 2.54 T. 
Figure 3 shows the recovery curves $m(t)$ of the nuclear magnetization $M(t)$ of La which are consistently fitted to the theoretical curve with a unique $T_1$\cite{Narath}. 
$1/T_1$ exhibits a divergence at $T_N =$142 K, and the overall temperature dependence of $1/T_1$ in Fig.~2 can be fitted to the SCR theory for weak itinerant antiferromagnets:
\begin{eqnarray}
\frac{1}{T_1}=\left\{ \begin{array}{ll}
aT+bT/\sqrt{T-T_N} & T>T_N \\
cT/M(T) & T<T_N, 
\end{array} \right.
\label{SCR}
\end{eqnarray}
where $a=0.005$ (sK)$^{-1}$, $b=0.13$ s$^{-1}$K$^{-1/2}$ and $c/M_0=$0.02 (sK)$^{-1}$ are fitting parameters, and $M(T)=M_0(1-T/T_N)^{0.15}$ is the AFM order parameter mentioned above. The first term in (\ref{SCR}) comes from usual Korringa relaxation expected in a metal. It is noteworthy that $1/T_1$ related to the $q$-averaged dynamical susceptibility shows a divergence at 142 K whereas a clear anomaly is not observed in the static susceptibility. 
These are characteristic of an itinerant antiferromagnet as seen in V$_3$Se$_4$\cite{Kitaoka}, in which the ordered $\bm{q}$-vector is far from $\bm{q} = 0$. 
Although the SCR expression roughly captures the behavior of $1/T_1$ in LaOFeAs, the sharp decrease of $1/T_1$ just below $T_N$ cannot be reproduced with the expression. 
This reflects a structural phase transition from the tetragonal ($P4/nmm$) to orthorhombic ($Cmma$) at $\sim165$ K observed in the same batch sample as ours \cite{Nomura}. 
In fact, the plot of $T_1T$ against $T$ shown in the inset of Fig.~2 suggests that the magnetic fluctuations changes around 160 K where the structural phase transition occurs. 
One of the plausible scenarios for the occurrence of the magnetic ordering below the structural phase transition is that the cylindrical Fermi surfaces are distorted, and thus the nesting between Fermi surfaces are enhanced. 
To check this scenario, it is desired to investigate the relation between the structural transition temperature and $T_N$ with the same technique. 
Quite recently, it has been reported that the neutron scattering measurements revealed a small ordered moment 0.4 $\mu_B$/Fe in LaOFeAs below 134 K, which is slightly lower than the lattice anomalous temperature 155 K\cite{Cruz}. Our La NMR results are consistent with this neutron-scattering measurement\cite{Mandrus}.

\begin{figure}[tb]
\begin{center}
\includegraphics[width=6.3cm]{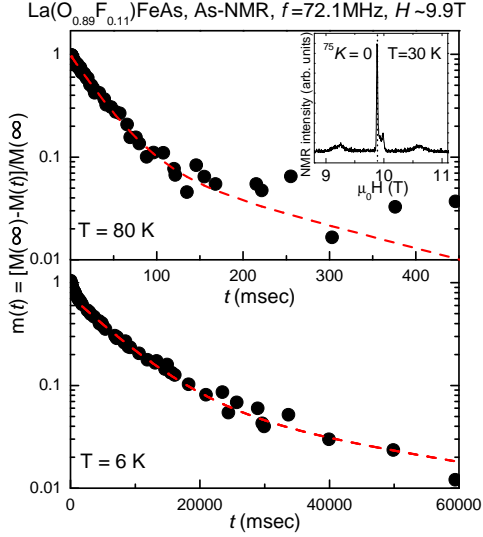}
\end{center}
\caption{The recoveries of the nuclear magnetization of As for $x=0.11$ fitted with $m(t)=0.1e^{(-t/T_1)}+0.9e^{(-6t/T_1)}$ (dotted lines), which were measured at the intense peak of the central transition as shown in the inset. The inset shows the As-NMR spectrum obtained by sweeping magnetic field at 72.1 MHz.}
\end{figure}
\begin{figure}[tb]
\begin{center}
\includegraphics[width=7.5cm]{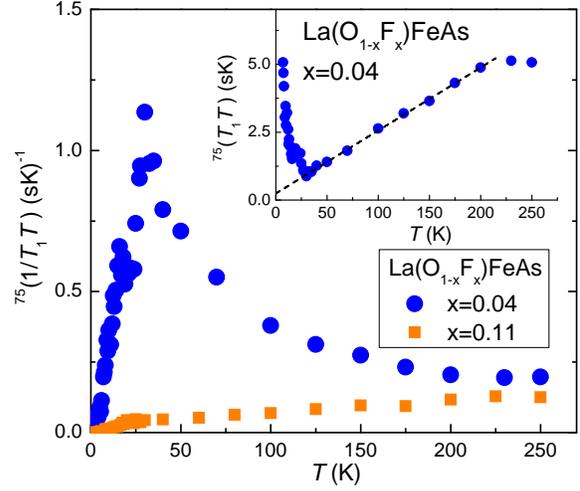}
\end{center}
\caption{ $T$-dependence of $(T_1T)^{-1}$ of $^{75}$As for $x=0.04$ and 0.11 in La(O$_{1-x}$F$_x$)FeAs. Inset: $T_1T$ as a function of $T$. The broken line is a fit to  $T_1T\propto T+\theta$ with $\theta=10.3\pm 2$ K.}
\end{figure}
Next, we turn to $^{75}$As-NMR results on the SC F-doped compounds.
Figure 5 shows the $T$-dependencies of $1/T_1T$ of As for $x=$0.04 and 0.11, in which $T_1$ was measured at the intense peak of the central transition obtained in the fixed frequency of 72.1 MHz (see the inset of Fig.~4). 
Figure~4 shows the recovery curves $m(t)$ at 80 and 6 K of 11\% F-doped sample. A single component of $T_1$ was consistently derived from $m(t)$ above 30 K, below which a short component of $T_1$ appears gradually.
The behavior of $1/T_1T$ for $x=0.04$ is strikingly different from that for $x=0.11$, suggesting that magnetic fluctuations strongly depend on the F-concentration.  
$1/T_1T$ for $x=0.04$ increases with decreasing temperature down to $\sim 30$ K. 
The $T$-dependence of $1/T_1T$ follows a Curie-Weiss law $1/T_1T \propto C / (T+\theta)$ between 30 and 200 K, which is clearly seen in the inset of Fig.~5. 
The obtained Weiss temperature is $\theta=10.3\pm2$ K, indicating that the $1/T_1T$ does not diverge at a finite temperature. This suggests that the superconductivity in the F-doped La(O$_{1-x}$F$_{x}$)FeAs emerges when a magnetic ordering is suppressed. 
This tendency is similar to that in $1/T_1T$ of Cu in underdoped La$_{2-x}$Sr$_x$CuO$_4$\cite{Ohsugi}, where superconductivity appears when a Weiss temperature obtained from $1/T_1T$ becomes positive. 

\begin{figure}[tb]
\begin{center}
\includegraphics[width=9cm]{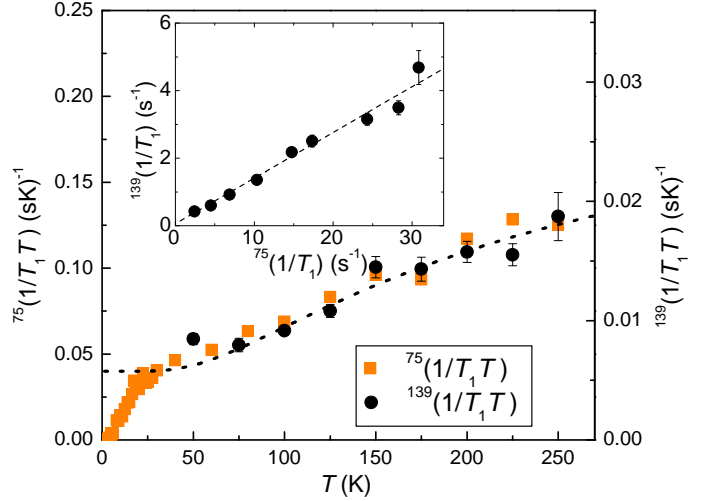}
\end{center}
\caption{$(T_1T)^{-1}$ for $x=$0.11 in La(O$_{1-x}$F$_x$)FeAs of $^{75}$As and $^{139}$La. The dashed line is a fit to $a+b\exp{(-\Delta_{\rm PG}/T)}$. Inset: $^{139}$($1/T_1$) at the La site plotted against $^{75}$($1/T_1$) at the As site.}
\end{figure}
Markedly different spin-dynamics from that for $x=0.04$ is observed for $x=0.11$, as shown in Fig.~6.
$1/T_1T$ both at the $^{75}$As and $^{139}$La site decreases with lowered temperature, which is reminiscent of the pseudogap behavior in underdoped regime of high-$T_c$ cuprates e.g. underdoped YBa$_2$Cu$_3$O$_{6.6}$\cite{Takigawa} and YBa$_2$Cu$_4$O$_8$\cite{Tomeno}, and approaches a nearly constant value in a narrow $T$-region just above $T_c$. 
The dashed line in Fig.~6 is a fit to 
\begin{equation}
\frac{1}{T_1T}=A+B\exp{(-\Delta_{\rm PG}/T)}, 
\end{equation}
where $A=$0.04 (sK)$^{-1}$, $B=0.17\pm0.01$ (sK)$^{-1}$, and $\Delta_{\rm PG}= 172\pm17$ K. 
Recent $^{19}$F NMR measurements also report pseudogap behavior for $x=0.11$\cite{Imai}. 
However, we point out a difference between the pseudogap behavior for $x=0.11$ and that in the underdoped cuprate. In the cuprate, $1/T_1T$ decreases from far above $T_c$ and no clear anomaly is observed at $T_c$. In contrast, for $x=0.11$, the Korringa behavior ($T_1T$ = const.) is observed in the narrow $T$-region from 30 K to $T_c$, which is related to the $T^2$ behavior of the resistivity, and the clear anomaly of $1/T_1T$ is found at $T_c$. These behaviors for $x=0.11$ are related to the multiband nature of the Fermi surfaces. We consider that the pseudogap originates from one of the multibands, and that the superconductivity is related with a band without showing the pseudogap. 
The inset shows the plot of $1/T_1$ of La against $1/T_1$ of As. A good linear relation indicates the spin dynamics at both sites are determined by the same fluctuations arising from the Fe-$3d$ spins. From the linear coefficient ($^{139}(1/T_1) \sim 0.135 \times^{75}(1/T_1)$), the ratio between the hyperfine coupling constant, ($^{139}H_{\rm hf}/^{75}H_{\rm hf}$), is estimated to be $\sim$ 0.45, showing that the coupling between two layers is not so small.    

\begin{figure}[tb]
\begin{center}
\includegraphics[width=6.5cm]{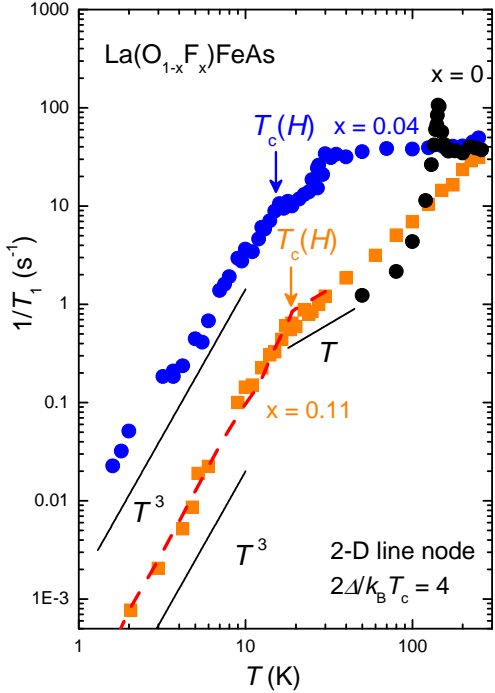}
\end{center}
\caption{$T$-dependence of $T^{-1}$ under $H \sim 9.9$ T for $x=$0.0, 0.04 ($T_c(H)=16 K$) and 0.11 ($T_c(H)=20 K$). The dashed curve is a calculation with assuming a line node gap $\Delta(\phi)=\Delta_0\sin{(2\phi)}$ with $\Delta_0=2k_BT_c$ (see text).}
\end{figure}
Next, we discuss the SC-gap properties from the $T$-dependence of $1/T_1$ in the SC state. Figure 7 shows the $T$-dependence of $1/T_1$ of As for $x=0.04$ and 0.11, together with $1/T_1$ of La in the undoped LaOFeAs. In this figure, $1/T_1$ of La is normalized by the relation ($^{139}(1/T_1)/^{75}(1/T_1) \sim$ 0.135). 
As noted above, a short component of $T_1$ appears gradually in both samples below 30 K and its fraction increases with decreasing $T$. Since the short component in $m(t)$ is about 30\% at 6 K for $x=0.11$, the main component (corresponding to the longer component) of $T_1$ is shown in Fig.~7.
We observed $T^3$ dependence of $1/T_1$ in the SC state for $x=0.04$ and 0.11, suggesting that the SC gap has line-nodes\cite{GangMuSpecificHeat, LeiShanPointContact, CongRenHc1, LuetkensmuSR, Imai}. We also found that $1/T_1$ decreases suddenly without showing Hebel-Slichter (coherence) peak. 
In general, a tiny coherence peak remains in anisotropic $s$-wave SC with a line-node gap ($\Delta(\phi)=|\Delta_0\sin{(2\phi)}|$)\cite{Borkowski}, since the coherence factor does not vanish when the gap function is integrated over the Fermi surfaces. The sharp decrease of $1/T_1$ just below $T_c$ and $T^3$ dependence in the SC state suggest that the superconductivity in La(O$_{1-x}$F$_{x}$)FeAs is classified into a non $s$-wave type. The observed $T^3$ dependence of $1/T_1$ data for $x=0.11$ can be reproduced using a 2-D line-node ($\Delta(\phi)=\Delta_0\sin{(2\phi)}$) model with $2\Delta/k_{\rm B}T_c = 4.0$. However, we point out that a residual density of states (DOS), which is suggested from the Korringa behavior at low temperatures, is not observed in both samples. In non $s$-wave superconductors, the residual DOS is usually induced by impurities and crystal imperfections. 
Since the absence of residual DOS appears to be contrary to non $s$-wave models, further NMR measurements using high quality samples are important in order to fully determine the SC symmetry. 

Finally, we would like to discuss the evolution of the spin dynamics in the normal state with F doping (corresponding to electron doping) and the relationship between magnetic fluctuations and superconductivity. 
From the Fermi-surface structure, it is considered that there exists two different kinds of fluctuations in this system, e.g. the fluctuations arising from the interband and intraband scatterings\cite{Kuroki}. 
The former gives rise to the AFM fluctuations and the latter induces the fluctuation near $\bm{q} = 0$. 
The AFM fluctuations observed in the undoped and 4\% F-doped samples are considered to originate from the interband fluctuations. 
Upon F doping, such AFM fluctuations disappear in the 11\% F-doped sample due to the shrinkage of the hole Fermi surfaces by the electron doping. 
Almost constant value of $T_c$ against the F concentration, irrelevant to the drastic change of $1/T_1T$ in the normal state, suggests that the superconductivity is related to the intraband fluctuations of the electron Fermi surfaces. 
For further insight on the relationship between magnetic fluctuations and superconductivity, $^{57}$Fe NMR measurements are needed.

In conclusion, the present NMR study revealed that an itinerant antiferromagnet in LaOFeAs evolves into superconductors with F doping. The spin fluctuations vary markedly with F doping, i.e. the strong AFM behavior for $x=0.04$ and the pseudogap behavior for $x=0.11$. 
The observed $T$-dependence of $1/T_1$ suggests unconventional superconductivity with a line-node SC gap in F-doped La(O$_{1-x}$F$_x$)FeAs.

We thank Y. Ihara, K. Kitagawa, Y. Maeno and K. Yoshimura for experimental supports and valuable discussions. We also thank H. Ikeda, S. Fujimoto and K. Yamada for valuable theoretical discussions. This work was supported by the Grant-in-Aid for the 21st Century COE "Center for Diversity and Universality in Physics" from the Ministry of Education, Culture, Sports, Science and Technology (MEXT) of Japan, and by the Grants-in-Aid for Scientific Research from Japan Society for Promotion of Science (JSPS). 
One of the authors (Y.N.) is financially supported as a JSPS Research Fellow.

\end{document}